\documentclass[twocolumn,showpacs,pre,floatfix,letter]{revtex4-1}

\ifx\pdfoutput\undefined
\usepackage{graphicx}
\else
\usepackage{epstopdf}
\usepackage{epsfig}
\usepackage{amssymb,amsmath,stmaryrd,tabularx}
\usepackage{wrapfig}
\usepackage{bm}
\fi

\usepackage[center]{subfigure}

\begin{document}
\title{Effect of weak rotation on large-scale circulation cessations in turbulent convection}
\author{Michael Assaf$^1$, Luiza Angheluta$^{1,2}$ and Nigel Goldenfeld$^1$}
\affiliation{$^1$Department of Physics, University of Illinois at
Urbana-Champaign, Loomis Laboratory of Physics, 1110 West Green
Street, Urbana, Illinois, 61801-3080\\
$^2$Physics of Geological Processes, Department of Physics, University of Oslo, Norway}


\begin{abstract}
We investigate the effect of weak rotation on the large-scale circulation
(LSC) of turbulent Rayleigh-B\'{e}nard convection, using the theory for
cessations in a low-dimensional stochastic model of the flow previously
studied. We determine the cessation frequency of the LSC as a function
of rotation, and calculate the statistics of the amplitude and
azimuthal velocity fluctuations of the LSC as a function of the
rotation rate for different Rayleigh numbers. Furthermore, we show
that the tails of the reorientation probability distribution function
remain unchanged for rotating systems, while the distribution of the
LSC amplitude and correspondingly the cessation frequency are strongly
affected by rotation. Our results are in close agreement with
experimental observations.
\end{abstract}
\maketitle

\def\Ra{\hbox{Ra}}
\def\Re{\hbox{Re}}
\def\Ro{\hbox{Ro}}

The complex phenomenon of thermal turbulence, also known as turbulent
Rayleigh-B\'{e}nard convection (RBC), develops in a heated fluid layer under
gravity by a succession of instabilities in the thermal transport due
to an interplay between different driving forces such as buoyancy,
viscous drag, and diffusion. This balance between different transport
mechanisms is quantified by the Rayleigh number $\Ra=\alpha_{_0}
g\Delta T L^3/\nu\kappa$ determining the flow state. Here,
$\alpha_{_0}$ is the isobaric thermal expansion coefficient, $g$ is the
gravity field, $\Delta T$ is the temperature gap between bottom and top
plates, $L$ is the height of the fluid container, $\kappa$ is the
thermal diffusivity and $\nu$ is the kinematic viscosity. When $\Ra$ is sufficiently large, the flow becomes turbulent and a large-scale
circulation (LSC) is formed. The latter is maintained by the emission of
plumes from the top and bottom surfaces which, due to buoyancy, move
upwards (hot plumes) or downwards (cold plumes)~\cite{Krishnamurti81,SIGG94,AHLE09}.
This LSC is known to appear in various \textit{rotating} natural systems, such as atmospheric~\cite{Vand00} and oceanic flows~\cite{MARS99}, and the dynamo driving planetary magnetic fields~\cite{ROBE00}.

Such a circulating state does not persist indefinitely, however, and
cessations followed by restarted flows at a new azimuthal angle occur
sporadically~\cite{Ahlers04,Ahlers06,AHLE09,Xi09,Ahlers10}. The complex
dynamics of the LSC is influenced by the heat and momentum transport
mechanisms, as well as by the geometry and aspect ratio $\Gamma$
(diameter over height) of the fluid container. In experiments using a
cylindrical geometry with $\Gamma=1$, the LSC occurs in a nearly
vertical plane, whereas for aspect ratio $\Gamma= 0.5$, two convection
rolls may coexist~\cite{Weiss11}. Recent studies also report that the
LSC flow-reversals and its tilted orientation are strongly influenced
by the corner-flows that form in a rectangular
geometry~\cite{SUGI2010}. The dynamics of a nearly-vertical, single LSC
can be modeled by a set of nonlinear stochastic differential equations
that describe the amplitude of azimuthal temperature variations,
$\delta$, and the azimuthal orientation angle,
$\theta_0$~\cite{Ahlers07,Ahlers08}. This model is found to be in
excellent agreement with the typical fluctuations of the
LSC~\cite{Ahlers07,Ahlers08}.  Yet, the cessations require an
extension of the model, since boundary terms describing the thermal
and viscous diffusion become dominant terms when the amplitudes are
small, as they inevitably must be during a cessation.  With this
extension, the model is able to provide an excellent description of the
cessation rates, as well as the LSC amplitude and  azimuthal velocity probability distribution functions (PDFs) including the
tails~\cite{AAG2011}.

The purpose of this Letter is to study the effect of weak rotation on
the statistics of the main degrees of freedom, \textit{i.e.} amplitude
and azimuthal velocity, and the LSC cessation events in
RBC.  Here the rotation rate $\Omega$ is conveniently represented by the dimensionless convective Rossby number,
$\Ro=(2\Omega)^{-1}\sqrt{\alpha_0 g\Delta T/L}$, which measures the
buoyancy relative to the Coriolis force. The total heat transport,
measured by the Nusselt number ${\cal N}$ relative to heat diffusion,
has a non-trivial dependence on $\Ro$, exhibiting three different
regimes~\cite{Ecke97,Ahlers09a,Ahlers10}: (i) an LSC-dominant regime at
$\Ro^{-1}\lesssim \Ro_c^{-1}\simeq 0.4$ where buoyancy dominates over
Coriolis effects, (ii) an Ekman-vortex pumping regime at
$\Ro_c^{-1}<\Ro^{-1}<\Ro_{max}^{-1}$ where the Coriolis force dominates over the buoyancy force for a non-vanishing $\Ro_c^{-1}$, and
(iii) at $\Ro^{-1}>\Ro_{max}^{-1}$ (typically $\Ro_{max}^{-1}\simeq
3-10$ depending on $\Ra$~\cite{Ahlers10}) where the velocity gradients,
and thus heat transport by convection, are suppressed via the
Taylor--Proudman effect~\cite{Ahlers10}. Here $\Ro_c^{-1}$ scales as
$1/\Gamma$ due to a finite-size
effect~\cite{Weiss2010finite,WeissHeat2011}, while $\Ro_{max}^{-1}$ is
determined by an interplay between the Ekman enhancement and the
Taylor--Proudman depression.

In this work we focus only on weak rotations, where the LSC is still present,
namely in the regime  $0\!<\!\Ro^{-1}\lesssim 0.6$. Recent experimental
studies report a non-monotonic dependence of the LSC mean amplitude on
$\Omega\sim \Ro^{-1}$~\cite{Ahlers10}. The mean amplitude initially
increases with increasing $\Ro^{-1}$ until $\Ro^{-1}$ becomes close to
(but below) a critical value, $\Ro_c^{-1}$. The origin of this increase
is not well understood, but may be associated with centripetal
effects~\cite{Stevens10}, see below. For higher $\Ro^{-1}$, the
detaching plumes from the thermal boundary layers (BLs) interact with
the Ekman-vortex structure that forms at $\Ro_c^{-1}$, so that these
plumes can no longer feed the LSC~\cite{Stevens10a}. Thus, the mean
amplitude decreases at $\Ro^{-1}\gtrsim\Ro_c^{-1}$. For
$\Ro^{-1}\gtrsim 0.6-1$ (depending on $\Ra$, $\Pr$ and $\Gamma$), the
mean amplitude becomes comparable to the root-mean-square temperature
fluctuations about the mean, and thus becomes
ill-defined~\cite{Ahlers10}. This non-monotonic behavior of the LSC
mean amplitude on $\Ro^{-1}$ is accompanied by a non-monotonic behavior
of the LSC cessation frequency, and this is what we calculate below.
This dependence contrasts with our result for the tails of the
reorientation statistics, namely the probability for a large angle
change in the LSC plane, which we show is almost independent of the
rotation strength in this weak rotation regime.

\medskip
\noindent {\it Statistics of the LSC dynamics with rotation:-\/} The
evolution of the dimensionless LSC amplitude, $\xi$, and azimuthal
angle, $\theta_0$, in the absence of rotations is governed by the
stochastic differential equations~\cite{AAG2011}:
\begin{eqnarray}\label{sdes}
\dot{\xi}&=&A+\alpha\xi-\beta\xi^{3/2}+f_{\xi}(t),\nonumber\\ \ddot
\theta_0  &=& -\left(\alpha_1\xi
+\beta_1\frac{\tau_{\dot\theta}}{\tau_\delta}\sqrt{\xi}\right)\dot\theta_0+f_{\dot\theta}(t).
\end{eqnarray}
Here, $A$ is a constant related to the heat transport or the inverse of
the thermal BL width, $\xi=\delta/\delta_0$ is the dimensionless LSC
amplitude, where $\delta$ is the physical LSC amplitude, $\delta_0
\approx {\Delta T\sigma}\Re^{3/2}/{\Ra}$ is the mean LSC amplitude, and
$\sigma=\nu/\kappa$ is the Prandtl number. Furthermore, the Reynolds
number is defined as $\Re=(\tau_{\delta}/\tau_{\dot\theta})^2\gg 1$
where $\tau_{\delta}$ and $\tau_{\dot\theta}$ are the turnover times in
the LSC and azimuthal planes, respectively, while time in
Eqs.~(\ref{sdes}) is measured in units of the corresponding turnover
times $\tau_{\delta}$ and $\tau_{\dot\theta}$~\cite{AAG2011}. Also,
delta-correlated Gaussian stochastic forcing terms $f_{\delta}(t)$ and
$f_{\dot\theta}(t)$ with amplitudes $D_{\delta}$ and $D_{\dot\theta}$
are included in Eqs.~(\ref{sdes}) to simulate the effect of turbulent
fluctuations. Finally, the coefficients
$\alpha,\beta,\alpha_1,\beta_1={\cal O}(1)$ are included to account for
the geometric coefficients from the spatial volume averaging
procedure~\cite{Ahlers07,Ahlers08}. Note, that in Eqs.~(\ref{sdes}) all
tildes (which appeared in~\cite{AAG2011} due to time rescaling) were
removed for clarity.

In the presence of rotation, the equation for the LSC amplitude [the
first of Eqs.~(\ref{sdes})] remains the same, since the same drag and
bouyancy forces drive the motion of the LSC in the vertical plane. Yet,
the coefficients in front of these terms may depend on $\Ro$, to
account for the experimental fact that the mean dimensionless LSC
amplitude $\xi_0$ is a function of $\Ro^{-1}$. To find this
$\Ro$-dependence, we notice that the coefficients $\alpha(\Ro)$,
$\beta(\Ro)$ and $A(\Ro)$ are related to each other by the constraints
that the PDF $P(\xi)$ is centered around $\xi=\xi_0$ and has a width
$D_{\delta}$~\cite{AAG2011}. Employing these constraints, we obtain
\begin{equation}\label{alphabeta}
\alpha(\Ro)=1-3A/\xi_0\;,\;\;\;\beta(\Ro)=\xi_0^{-1/2}\left(1-2A/\xi_0\right).
\end{equation}
Now we formulate a simple  theory that can explain the dependency
of these coefficients as well as that of $\xi_0$ on $\Omega$. We assume that
$\alpha=\alpha_0$, is constant (because the geometrical coefficient in
the buoyancy should not be affected by rotation) and expand
$\beta=\beta_0(1+b_1\Omega+b_2\Omega^2)$ (assuming that the thickness
of the viscous BL is dependent on $\Omega$), where $\alpha_0=1-3A_0$,
$\beta_0=1-2A_0$ and $A_0=A(\Omega=0)$. In this way, we can find
expressions for $A(\Omega)$ and $\xi_0(\Omega)$, by equating these
expressions for $\alpha$ and $\beta$ with Eq.~(\ref{alphabeta}), and keeping terms up to ${\cal
O}(\Omega^2)$. We obtain $\xi_0(\Omega)\simeq
1-2b_1\Omega + (3b_1^2-2b_2)\Omega^2$ and $A(\Omega)=A_0\xi_0(\Omega)$.
Note, that plugging this into Eq.~(\ref{alphabeta}), we obtain
$\alpha(\Ro)=\alpha_0$ and $\beta(\Ro)\simeq \beta_0\xi_0^{-1/2}$.
Because $\beta(\Ro)$ represents the inverse Rossby-dependent width of
the viscous BL, the latter increases as $\xi_0^{1/2}$ for
$0<\Ro^{-1}<\Ro_c^{-1}$, in agreement with recent experimental
observations~\cite{Stevens10a,Stevens10}. The origin of the increase in
the viscous BL width in the weak-rotation regime can be associated with
centripetal effects~\cite{Stevens10} that tend to increase the BL width
according to the Prandtl-Blasius theory.

\begin{figure}
\includegraphics[width=0.95\columnwidth]{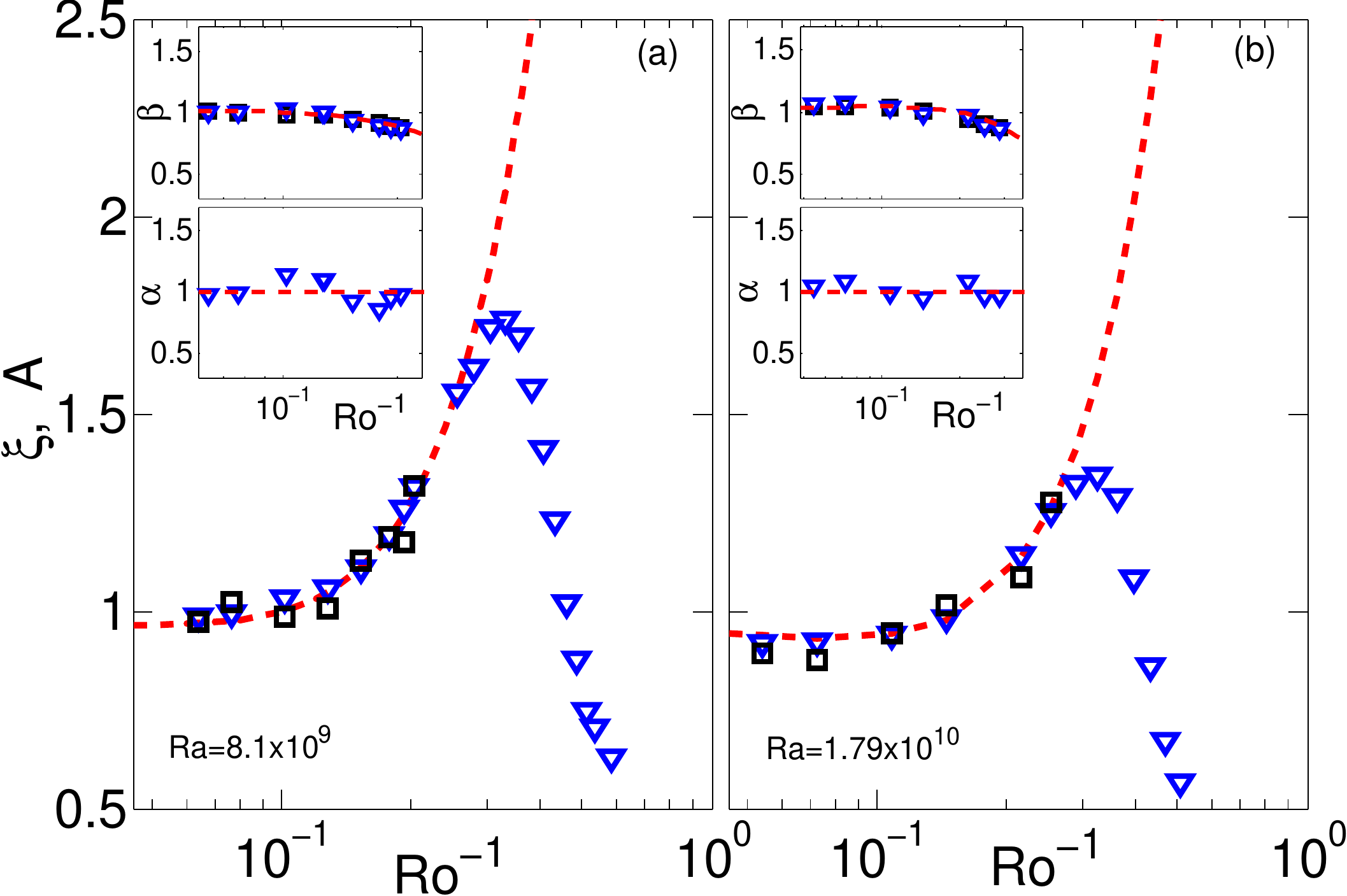}
\caption{(Color online). Shown are the values of $\xi_0$  (triangles),
and $A$ (squares). The dashed line is a parabolic fit. The upper inset
shows $\beta$ (triangles) compared with $\xi_0^{-1/2}$ (squares), while
the dashed line is a parabolic fit. The lower inset shows $\alpha$ and
the dashed line is a guide for the eye. All quantities are normalized
by their non-rotating values.}\label{A}
\end{figure}

In Fig.~\ref{A} we plot the experimental mean amplitude $\xi_0$ and $A$
(normalized by their value at zero rotation) as a function of
$\Ro^{-1}$. According to our theory, provided that $\xi_0$ is
well-approximated by a parabolic expansion in $\Omega$, in the weak
rotation regime, $A/A_0$ should coincide with $\xi_0/\xi_0(0)$, and
this is indeed the case. The insets in this figure confirm the
dependencies of $\alpha$ and $\beta$ on $\Omega$ for two different
$\Ra$ numbers. Note that in order to extract $A$ from the experimental
data, we have used the relation $A=BD_{\delta}/2$~\cite{AAG2011} where
$B$ and $D_{\delta}$ are specified below.

\begin{figure}
\includegraphics[width=.950\columnwidth]{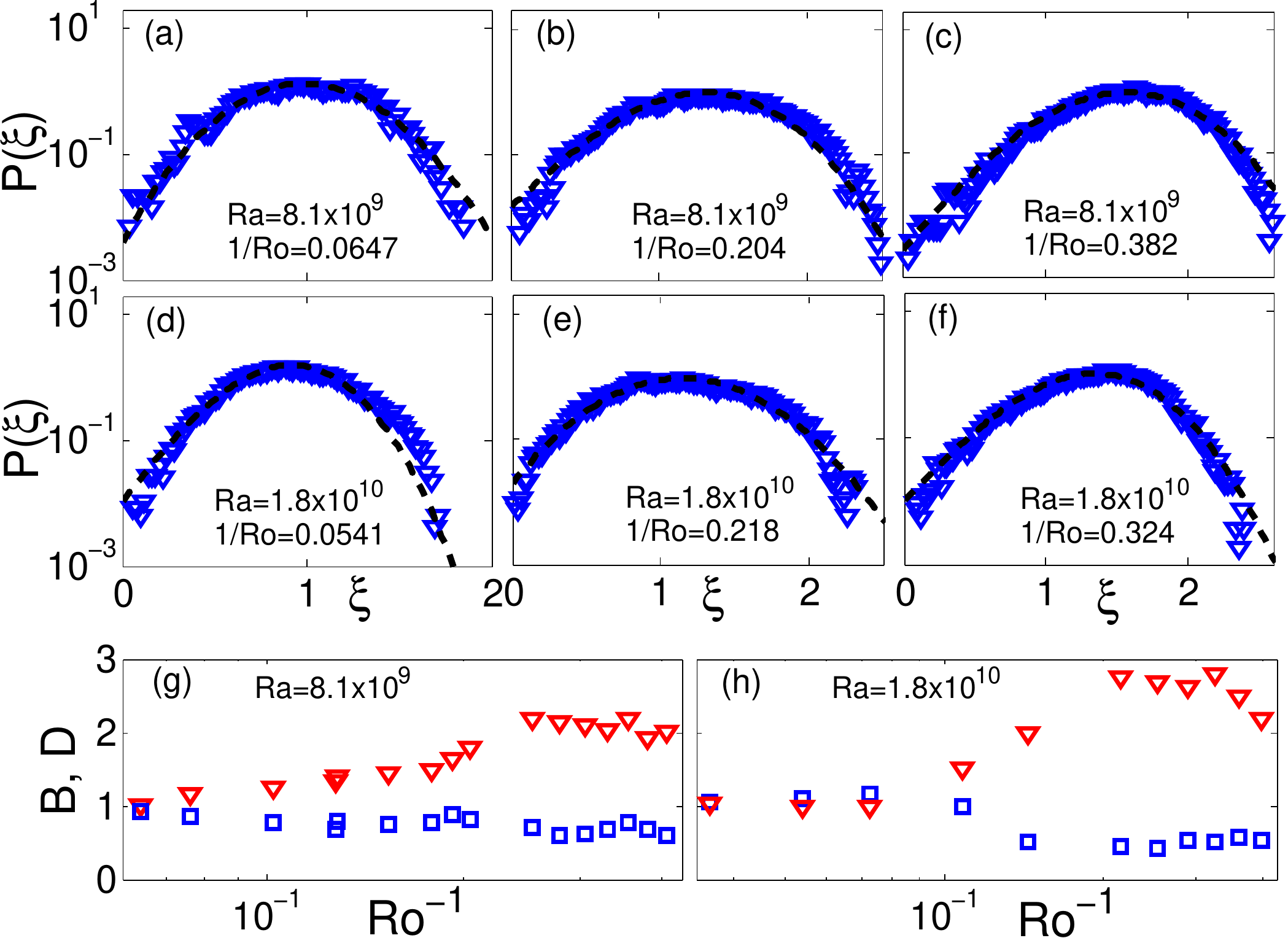}
\caption{(Color online). In panels (a)-(f) shown are PDFs of the LSC
rescaled amplitude $\xi$ for different $\Ro$ and $\Ra$ numbers. Each
row represents a different $\Ra$ number. The experimental PDFs are
represented by triangles, while the fitting curves (dashed line) are
the analytical PDFs~(\ref{pdfxi}) with parameters $B$ and $D_{\delta}$
determined experimentally for each $\Ro$ number. The width of the PDFs
initially increases with increasing $\Ro^{-1}$ while the slope of the
left tail decreases. This can be seen in panels (g) and (h)
where we plot the experimental $B$ and $D_\delta$ as functions of
$\Ro^{-1}$ for different $Ra$.} \label{fig_PDFs}
\end{figure}

Since the equation for the LSC amplitude [the first of
Eqs.~(\ref{sdes})] is independent of the azimuthal reorientation angle
$\theta_0$, we can analyse it separately by writing the corresponding
Fokker-Planck equation and finding its stationary
solution~\cite{AAG2011}. Defining the potential
\begin{equation}
V(\xi) \!=\! -\frac{B D_\delta}{2}\xi\!-\!\left(1\!-\!\frac{3B
D_\delta}{2\xi_0}\right)\!\frac{\xi^2}{2}+\frac{2}{5{\xi_0}^{1/2}}\left(1\!-\!\frac{B
D_\delta}{\xi_0}\right)\!\xi^{5/2}
\end{equation}
the frequency-dependent PDF, $P(\xi,\Ro)$ takes the form
\begin{eqnarray}\label{pdfxi}
P(\xi,\Ro)=Ce^{-2V(\xi)/D_{\delta}}.
\end{eqnarray}
Here $C$ is a normalization constant, $D_{\delta}=D_{\delta}(\Ro)$ is
the PDF width in the Gaussian regime: $P(\xi\approx \xi_0)\sim
e^{-(\xi-\xi_0)^2/(2 D_{\delta})}$, and $B=B(\Ro)$ is the logarithmic
derivative of the PDF at small $\xi$, since $P(\xi\ll 1)\sim e^{B\xi}$.

Furthermore, it has been shown that the cessation frequency -- the
frequency of events that the LSC amplitude goes below a threshold
amplitude $\xi_{min}\ll 1$ -- is given by
\begin{equation}\label{eq:omega}
\omega^{-1}\!=\!\frac{1}{\xi_{min}}\int_0^{\xi_{min}}\!\!d\xi^*T(\xi^*)\,;\;\;T(\xi^*)\sim
e^{-2D^{-1}[V(\xi^*)-V(1)]}.
\end{equation}
Here $T(\xi^*)$ is the mean time it takes the amplitude to reach  $\xi^*\ll 1$.

In order to compare the results for the PDF [Eq.~(\ref{pdfxi})] and
cessation frequency [Eq.~(\ref{eq:omega})] with experimental results,
we extract the values of $B(\Ro)$ and $D_{\delta}(\Ro)$ from the
experimental PDFs, just as was done for the non-rotating
case~\cite{AAG2011}. In Fig.~\ref{fig_PDFs}, we compare experimental
and theoretical PDFs for different $\Ro$ and $\Ra$ numbers, in the
weakly-rotating regime. The theoretical predictions hold well for
various rotation frequencies, provided that we use the corresponding
frequency-dependent parameters. As $\Ro^{-1}\sim \Omega$ is increased,
the width of the PDF increases compared to the non-rotating case while
the slope of the left tail decreases. This functional dependence of the
experimental $B$ and $D_{\delta}$ on $\Ro^{-1}$ is shown in the lower
panels of Fig.~\ref{fig_PDFs} for two different $\Ra$ number.

\begin{figure}
\includegraphics[width=.80\columnwidth]{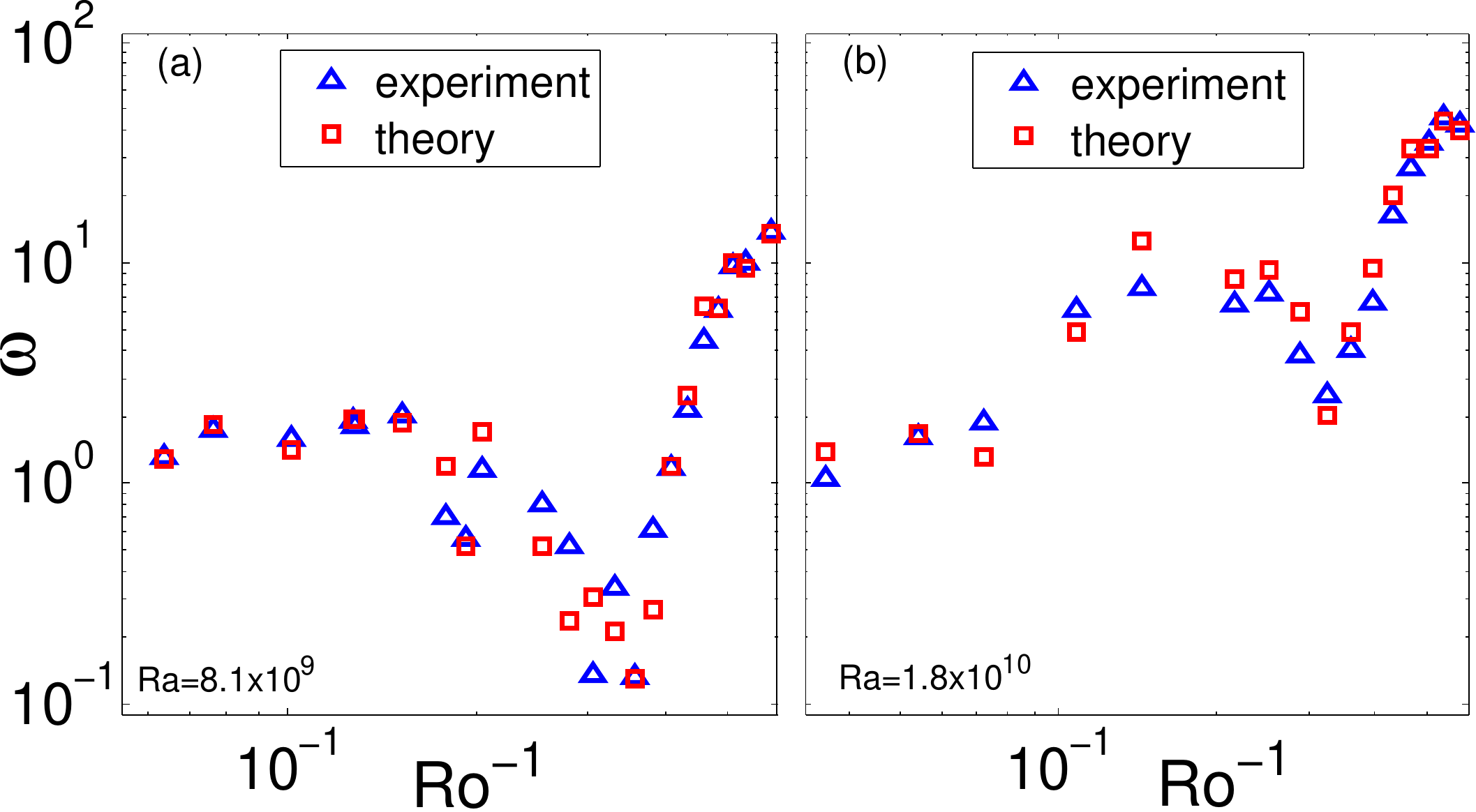}
\caption{(Color online). Cessation frequency (normalized by its
non-rotating value) as a function of $\Ro^{-1}$ for different values of
$\Ra$ number. The triangles are experimental results, while the squares
are the theoretical predictions according to Eq.~(\ref{eq:omega}). The
cessation threshold chosen here was $0.15$.} \label{fig_omega}
\end{figure}

Finally, in Fig.~\ref{fig_omega} we compare the theoretical and
experimental results for the cessation frequency as function of $\Ro$
for different $\Ra$ values and observe good agreement. Here, the
threshold for cessation was chosen to be $0.15$.

\medskip
\noindent
{\it Rotation rate of the azimuthal plane:-\/} In this section we investigate the effect of weak rotation on the
PDF for the angular velocity of the
azimuthal plane [the second of Eqs.~(\ref{sdes})]. Rotation brings
about a Coriolis force which is proportional to
$\vec{\Omega}\times\vec{V}_{\theta}$, where $|\vec{V}_{\theta}|\sim
\xi$~\cite{Ahlers07}. As a result, the effect of rotation on the
azimuthal dynamics amounts to the addition of a term of the form
$\xi\Omega \Phi(\theta_0)$ to the equation for $\dot{\theta}_0$, where
$\Phi={\cal O}(1)$ is some function of the azimuthal
angle~\cite{Ahlers06a}. Therefore, the second of Eqs.~(\ref{sdes})
under rotation becomes
\begin{equation}\label{thetaeq}
\ddot \theta_0  = \xi\Omega\Phi(\theta_0)-\left(\alpha_1\xi
+\beta_1\frac{\tau_{\dot\theta}}{\tau_\delta}\sqrt{\xi}\right)\dot\theta_0+f_{\dot\theta}(t).
\end{equation}
As expected, the addition of the first term on the right-hand-side of
Eq.~(\ref{thetaeq}) changes the steady state solution for
$\dot{\theta}$ at $\xi\simeq 1$ from $\dot{\theta}=0$ at zero rotation,
to $\dot{\theta}\sim\Omega$ at $\Omega>0$ [since $\Phi(\theta_0)={\cal
O}(1)$ and $\tau_{\dot\theta}/\tau_\delta\ll 1$]. This has been
experimentally observed by Zhong and Ahlers~\cite{Ahlers10} in the
weakly-rotating regime, below $\Ro_c^{-1}$. The reason for this
limitation is that the steady-state solution is valid only as long as
the parameters involved are of order unity, which holds for not too
large $\Omega$, but breaks down for high rotation frequencies. In the
latter case, when Ekman vortices start to form, the model becomes
invalid.

\begin{figure}
\includegraphics[width=.82\columnwidth]{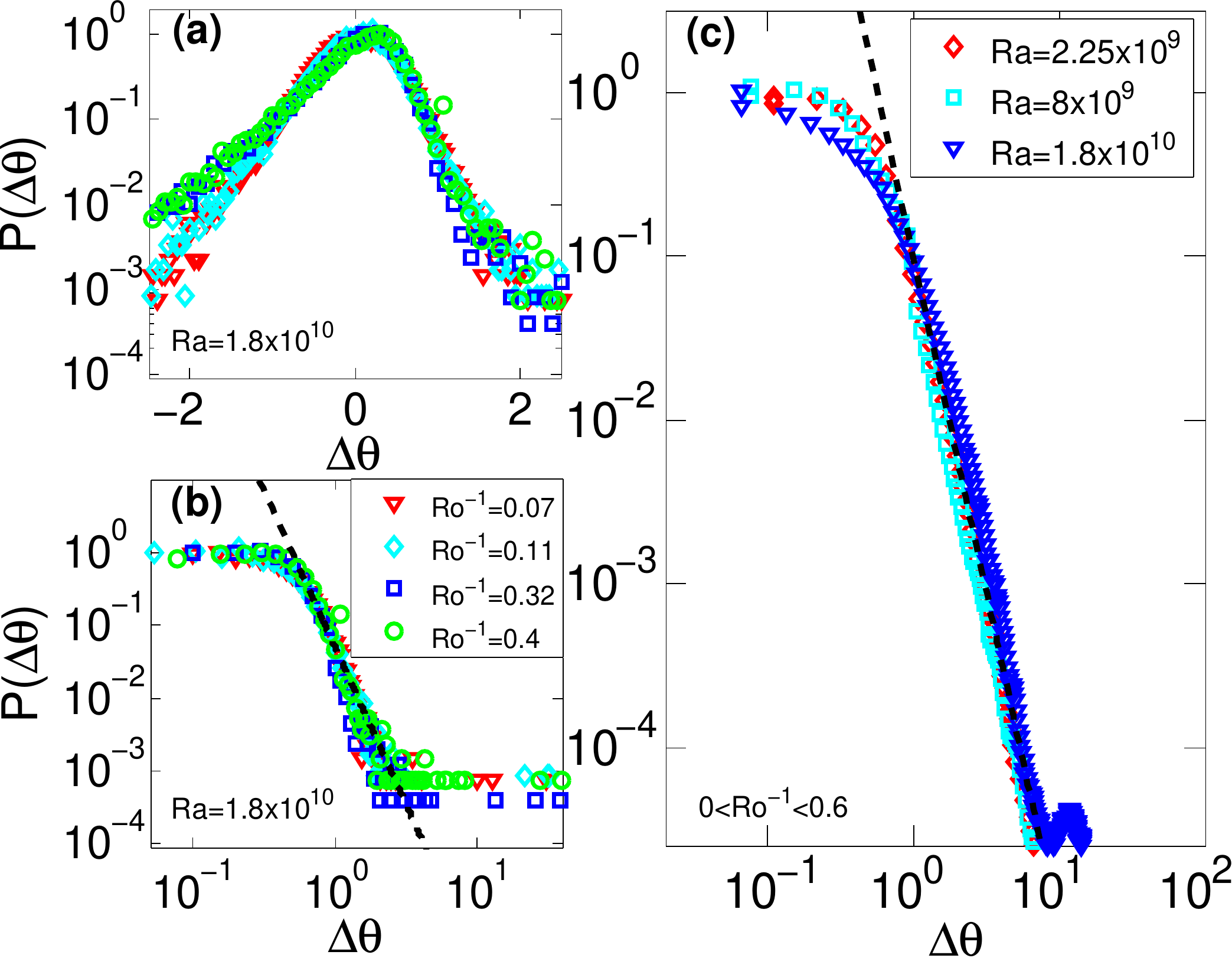}
\caption{(Color online). PDF $P(\Delta\theta)$ as function of
$\Delta\theta$. In (a) shown are experimental PDFs for four different
$\Ro$ numbers (see legend in panel b) for $\Ra=1.8\times 10^{10}$. As
$\Ro^{-1}\sim\Omega\,$ increases, the left tail becomes fatter while
the right tail remains more or less constant. In (b) shown are the same
PDFs on a log-log plot, demonstrating that the tail is well described
by a power law independent of $\Ro$; the dashed line is a power law
with exponent $-4.3$. In (c) shown are PDFs, averaged over a wide range
of $\Ro^{-1}$ numbers (generally $0<\Ro^{-1}<0.6$), for different
$\Ra$. The dashed line is the same as in panel (b). This panel implies
that the rare-event statistics of $P(\Delta\theta)$ are universal and
independent of $\Ra$ and $\Ro$. } \label{PDFs_theta}
\end{figure}

Apart from the steady-state solution that represents the mean angular
velocity, here we are mainly interested in the rare events of large
deviations in $\Delta\theta\sim\dot{\theta}$. The tails of the PDF
$P(\Delta\theta)$ were calculated in~\cite{AAG2011} for non-rotating
systems. Here, we show that these tails are unchanged by the rotation.
The reason is that the term due to rotation added to
Eq.~(\ref{thetaeq}) is proportional to $\xi$. As a result, large
deviations in $\Delta\theta$, which occur when $\xi\ll 1$, are still
governed by the term proportional to $\sqrt{\xi}\dot\theta_0$ in
Eq.~(\ref{thetaeq}), which is dominant in the regime of $\xi\ll
1$~\cite{AAG2011}. This term does not depend on the rotation frequency.
Therefore, the tails of $P(\Delta\theta)$ remain unchanged when
rotation is introduced, as shown in Fig.~\ref{PDFs_theta}.

Indeed, in the non-rotating case the tails have been found to scale as
a power law with exponent $-4$, while experimental results for
non-rotating systems demonstrate a power-law tail with exponent $\sim
-4.3$~\cite{AAG2011}. In Fig.~\ref{PDFs_theta}(b) shown are
experimental PDFs, $P(\Delta\theta)$, for different $\Ro$ numbers. In
panel (c) we show PDFs \textit{averaged} over a wide range of
$\Ro^{-1}$ numbers~\cite{Ahlers10} (generally $0<\Ro^{-1}<0.6$), for
different $\Ra$ numbers. The dashed lines are the theoretical
prediction: a power law curve with exponent $-4.3$. The excellent
agreement between theory and experiment indicates that the exponent for
rotating systems remains unchanged compared to the non-rotating case.
Note, that panels (b) and (c) are shown on a log-log scale which only
allows us to show the positive $\dot{\theta}$ region. We have checked that
the left and right tails of the PDF scale with the same exponent within
$5\%$.

\medskip
\noindent {\it Angular-dependent asymmetry of the PDF for azimuthal
fluctuations:-\/} We conclude by pointing out an interesting corollary
of our analysis.  From Fig.~\ref{PDFs_theta}(a), it is apparent that in
contrast to the non-rotating case, the PDF $P(\Delta\theta)$ develops
an asymmetric shape as a function of $\Delta\theta$ that is an
increasing function of the rotation rate, up to the vicinity of
$1/\Ro\sim 1/\Ro_c$.  We speculate that this asymmetry in the azimuthal
velocity fluctuations is related to spiral defects and their preferred
motion. Near the onset of convection, spiral defects are formed in the
presence of rotation and their azimuthal motion is \textit{against} the
direction of rotation~\cite{Zhong93}. With increasing Ra number, the
flow becomes more turbulent and the effect of the individual vortices
on the azimuthal velocity statistics diminishes~\cite{Vorobieff02}.
However, as the strength of the LSC decreases when $\Ro^{-1}$ exceeds
$\Ro_c^{-1}$, the flow, hence the velocity statistics, is more
influenced by the preferred motion of the spiral defects in the
presence of rotation, which would explain the observed broadening of
the left-tail of the PDF $P(\Delta\theta)$ corresponding to high
fluctuations against the direction of rotation. This effect of the
left-tail broadening with increasing $\Omega$ can be clearly seen in
Fig.~\ref{PDFs_theta}(a). However, the mechanism relating the asymmetry
in  $P(\Delta\theta)$ to spiral defects needs further elaboration, and
is beyond the scope of this paper.

We thank J.-Q. Zhong and G. Ahlers for sharing with us their
unpublished data on rotating RBC, shown in the figures.  M.\,A.
gratefully acknowledges the Center for the Physics of Living Cells at
the University of Illinois for support. L.\,A. is grateful for support
from the Center of Excellence for Physics of Geological Processes. This
work was partially supported by the National Science Foundation through
grant number NSF-DMR-1044901.
\bibliographystyle{apsrev4-1}
\bibliography{references}

\end{document}